\documentclass[prb,twocolumn,showpacs,preprintnumbers,amsmath,amssymb,superscriptaddress, bibnotes]{revtex4-1}

\usepackage{color}
\usepackage{graphicx}
\usepackage{dcolumn}
\usepackage{bm}

\begin{document}

\title{$S$ wave superconductivity in newly discovered superconductor BaTi$_2$Sb$_2$O revealed by $^{121/123}$Sb-NMR/Nuclear Quadrupole Resonance measurements}

\author{S.~Kitagawa}
\email{shunsaku@scphys.kyoto-u.ac.jp}
\author{K.~Ishida}
\affiliation{Department of Physics, Graduate School of Science, Kyoto University, Kyoto 606-8502, Japan}

\author{K.~Nakano}
\author{T.~Yajima}
\author{H.~Kageyama}
\affiliation{Department of Energy and Hydrocarbon Chemistry, Graduate School of Engineering, Kyoto University, Kyoto 615-8510, Japan}

\date{\today}

\begin{abstract}
We report the $^{121/123}$Sb-NMR/nuclear quadrupole resonance (NQR) measurements on the newly-discovered superconductor BaTi$_2$Sb$_2$O with a two-dimensional Ti$_2$O square-net layer formed with Ti$^{3+}$ (3$d^1$).
NQR measurements revealed that the in-plane four-fold symmetry is broken at the Sb site below $T_{\rm A} \sim$ 40 K, without an internal field appearing at the Sb site. These exclude a spin-density wave (SDW)/ charge density wave (CDW) ordering with incommensurate correlations, but can be understood with the commensurate CDW ordering at $T_{\rm A}$. 
The spin-lattice relaxation rate $1/T_1$, measured at the four-fold symmetry breaking site, decreases below superconducting (SC) transition temperature $T_{\rm c}$, indicative of the microscopic coexistence of superconductivity and the CDW/SDW phase below $T_{\rm A}$.
Furthermore, $1/T_1$ of $^{121}$Sb-NQR  shows a coherence peak just below $T_{\rm c}$ and decreases exponentially at low temperatures. 
These results are in sharp contrast with those in cuprate and iron-based superconductors, and strongly suggest that its SC symmetry is classified to an ordinary $s$-wave state.
\end{abstract}

\pacs{76.60.-k,	
71.20.Be	
74.25.-q 
75.25.Dk	
}

\abovecaptionskip=-5pt
\belowcaptionskip=-10pt

\maketitle

After the discovery of high-$T_{\rm c}$ cuprate superconductors, much efforts have been paid to synthesize new high-$T_{\rm c}$ superconductors. These activities have brought us the discovery of various unconventional superconductors, (e.g. Sr$_2$RuO$_4$\cite{Y.Maeno_Nature_1994,A.Mackenzie_RMP_2003,Y.Maeno_JPSJ_2012}, NaCoO$_2\cdot$1.5H$_2$O\cite{K.Takada_Nature_2003}, and LaFeAs(O$_{1-x}$F$_x$)\cite{Y.Kamihara_JACS_2008,K.Ishida_JPSJ_2009}) so far. 
It is quite interesting that all these superconductors possess a two-dimensional layered structure and are located near the magnetic instability, both of which are regarded as essential ingredients of high-$T_{\rm c}$ superconductors. 
Actually $T_{\rm c}$ in a quasi-two dimensional system is theoretically shown to be higher than in a three-dimensional system, since magnetic fluctuation is generally enhanced by low-dimensionality\cite{Y.Yanase_PhysRep_2003}.  
          
Quite recently, it was reported that two-dimensional oxyantimonide  BaTi$_2$Sb$_2$O shows a superconducting (SC) transition at $T_{\rm c} \sim 1$ K\cite{T.Yajima_JPSJ_2012,P.Doan_JACS_2012}. 
BaTi$_2$Sb$_2$O possesses a similar crystal structure as cuprate La$_2$CuO$_4$ and has a Ti$_2$O square net, which is an anti-configuration to the CuO$_2$ square net, as shown in Fig.~1. 
\begin{figure}[tb]
\vspace*{-10pt}
\begin{center}
\includegraphics[width=8cm,clip]{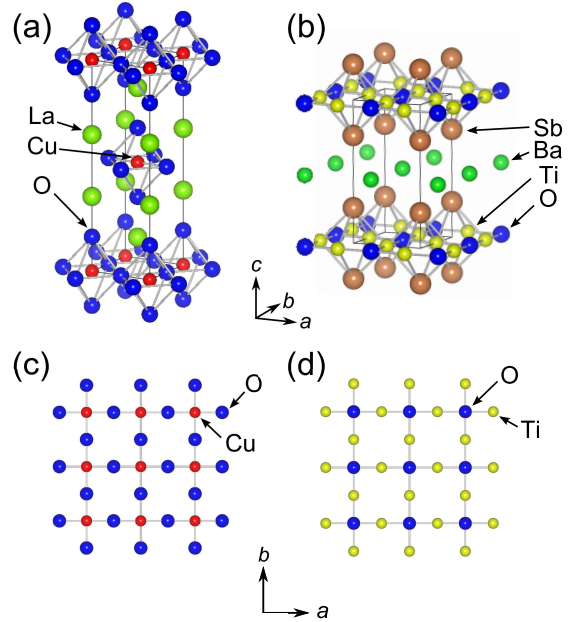}
\end{center}
\caption{(Color online)The crystal structure of (a) La$_2$CuO$_4$ and (b) BaTi$_2$Sb$_2$O.
Both compounds possess a layer structure. (c) Two dimensional CuO$_2$ plane in La$_2$CuO$_4$, and (d) two dimensional Ti$_2$O plane in BaTi$_2$Sb$_2$O.}
\label{Fig.1}
\end{figure}
The edge-shared TiO$_2$Sb$_4$ octahedra form the square lattice, and the electronic state of Ti$^{3+}$ is in the 3$d^1$ state, which is regarded as an electron-hole symmetric state of 3$d^9$ state in Cu$^{2+}$\cite{T.Yajima_JPSJ_2012}. 
Thus, BaTi$_2$Sb$_2$O is an interesting reference compound of cuprate superconductors although BaTi$_2$Sb$_2$O is metallic.
From the measurements of magnetic susceptibility and electrical resistivity, an anomaly was found at 50~K, and the occurrence of a charge density wave (CDW) or spin density wave (SDW) transition was suggested\cite{T.Yajima_JPSJ_2012,P.Doan_JACS_2012}.
Similar anomaly was observed in $A$Ti$_2$$Pn_2$O [$A$ = Na$_2$, Ba, (SrF)$_2$, (SmO)$_2$; $Pn$ = As, Sb] which have almost the same structure as BaTi$_2$Sb$_2$O and show no superconductivity\cite{E.A.Axtell_JSSC_1997,X.F.Wang_JPhys_2010,R.H.Liu_CM_2010}
, however details of this anomaly have not been investigated. 
Thus, the understanding of this anomaly, the electronic state of the Ti$^{3+}$ (3$d^1$) and the SC symmetry are important for seeking another way for high-$T_{\rm c}$ superconductivity in the strong correlated 3$d$ electron systems. 
In order to address above underlying issues, we performed $^{121/123}$Sb-NMR/nuclear quadrupole resonance (NQR) measurements on BaTi$_2$Sb$_2$O.

BaTi$_2$Sb$_2$O was synthesized by the conventional solid state reaction method\cite{T.Yajima_JPSJ_2012}.
NMR measurements are performed in the same batch as magnetic susceptibility and resistivity measurements\cite{T.Yajima_JPSJ_2012}.
To prevent sample degradation by air and/or moisture, poly-crystalline sample was mixed with stycast 1266, and the mixture was solidified with random crystal orientation. All procedures were done in a glove box filled with N$_2$.
$T_{\rm c}$ $\simeq$ 0.95~K of the sample was determined from the diamagnetic shielding signal, which is consistent with the previous report\cite{T.Yajima_JPSJ_2012}. No reaction was recognized between the sample and stycast, since $T_{\rm c}$ and $T_{\rm A}$ are unchanged during our measurements.

\begin{figure}[tb]
\vspace*{-15pt}
\begin{center}
\includegraphics[width=9cm,clip]{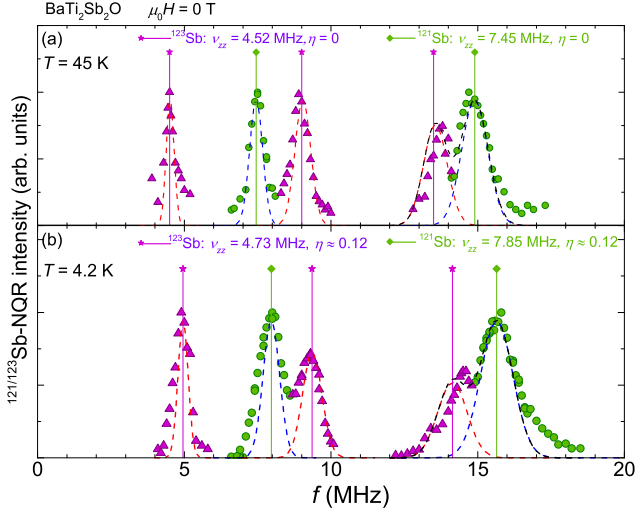}
\end{center}
\caption{(Color online) $^{121/123}$Sb-NQR spectra, which were obtained by frequency-swept method at (a) 45~K and (b) 4.2~K.
From the observed $^{121/123}$Sb-NQR spectra, the quadrupole parameters for the Sb nuclei were evaluated as shown in the figure.
The broken curves are the simulation of NQR spectra using the estimated quadrupole parameters.
The asymmetry parameter $\eta$ becomes finite at 4.2~K, indicative of the in-plane four-fold symmetry breaking at Sb site.
Line widths of NQR spectra can be explained by only the electric field gradient distribution, indicative of absence of internal magnetic field at Sb site.
}
\label{Fig.2}
\vspace*{15pt}
\end{figure}

\begin{table}[tp]
\caption[]{The data of Sb isotopes; the nuclear gyromagnetic ratio: $\gamma_n$, the nuclear quadrupolar moment: $Q$, natural abundance: N.A. and the nuclear spin: $I$. } 
\label{table:Sb_parameter}
\vspace{1cm}
\begin{tabular}{rccccc}
\hline
                   & $\gamma_n/2\pi$(MHz/T) & $Q$(10$^{-24}$cm$^2$ ) & N.A.(\%)  &  $I$\\ \hline
$^{121}$Sb & 10.189 & $-0.2 \sim -1.8$ & 57.3 & 5/2 \\
$^{123}$Sb & 5.5175 & $-0.2 \sim -0.7$ & 42.7 & 7/2  \\
\hline
\end{tabular}
\label{Tab.1}
\end{table}


Figure~\ref{Fig.2} shows the $^{121/123}$Sb-NQR spectra, which were obtained by frequency-swept method at 45~K ($> T_{\rm A}$) and 4.2~K ($< T_{\rm A}$).
There are two isotopes of Sb nuclei, properties of which are summarized in Table I. 
When $I \ge 1$, a nucleus has an electric quadrupole moment $Q$ as well as a magnetic dipole moment, and thus the degeneracy of nuclear-energy levels is lifted even at zero magnetic field due to the interaction between $Q$ and the electric field gradient (EFG). This interaction is described as 
\begin{align}
\mathcal{H}_Q &= \frac{\nu_{zz}}{6}\left\{(3I_z^2-I^2)+\frac{1}{2}\eta(I_+^2+I_-^2)\right\},
\label{eq.1}
\end{align}
where $\nu_{zz}$ is the quadrupole frequency along the principal axis ($c$-axis) of the EFG, and is defined as $\nu_{\rm zz} \equiv 3e^2qQ/2I(2I-1)$ with $eq = V_{zz}$,  and $\eta$ is an asymmetry parameter of the EFG expressed as $(V_{xx}-V_{yy})/ V_{zz}$ with $V_{\alpha \alpha}$ which is the EFG along $\alpha$ direction ($\alpha = x,y,z,$).
When $^{121}$Sb ($^{123}$Sb) is in the presence of the EFG, the degenerate six (eight) nuclear-spin states are split into three (four) energy levels, yielding two (three) resonance frequencies as shown in Fig.~\ref{Fig.2}.
The quadrupole parameters $\nu_{zz}$ and $\eta$ for each Sb nuclei are estimated from the comparison between the observed $^{121/123}$Sb-NQR spectra and calculated resonance frequencies obtained from the diagonalization of eq.\eqref{eq.1} as shown in Fig.~\ref{Fig.2}.
Field-swept NMR spectra can be consistently fit by the simulation calculated with the same quadrupole parameters as shown in Fig.~\ref{Fig.4}, indicative of the validity of the NQR analysis.
\begin{figure}[tb]
\vspace*{-10pt}
\begin{center}
\includegraphics[width=9cm,clip]{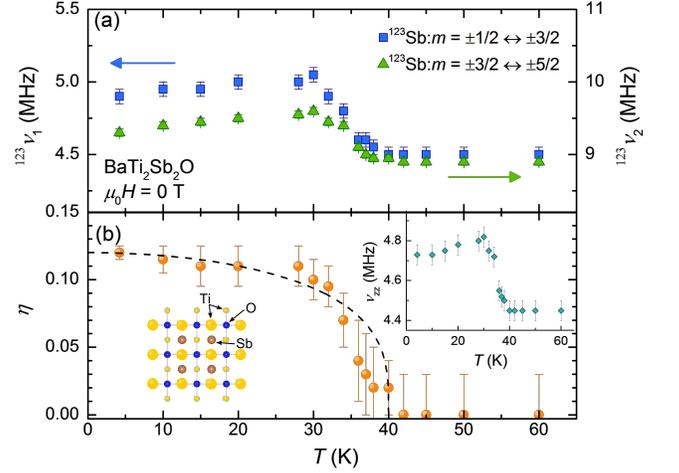}
\end{center}
\caption{(Color online) $T$ dependence of (a) resonance frequencies of $\pm$1/2 $\leftrightarrow$ $\pm$3/2 ($\nu_1$) and $\pm$3/2 $\leftrightarrow$ $\pm$5/2 ($\nu_2$) transition of $^{123}$Sb and (b) $\eta$. 
The quadrupole parameters $\nu_{zz}$ and $\eta$ are estimated from these resonance frequencies. 
The $\eta$ changes continuously below $\simeq 40$~K and no appreciable hysteresis is observed, indicating that this phase transition is 2nd order. The broken line is guide to eyes. 
The schematic image of the CDW state which is one of promising states below $T_{\rm A}$ is illustrated, where the difference of the circle size at the Ti site represents the difference of the charge densities of Ti.
(Inset) $T$ dependence of $\nu_{zz}$. 
The variation in $\nu_{zz}$ is consistent with that in lattice parameters, indicative of the validity of the NQR analysis.}
\label{Fig.3}
\end{figure}

\begin{figure}[tb]
\vspace*{-10pt}
\begin{center}
\includegraphics[width=9cm,clip]{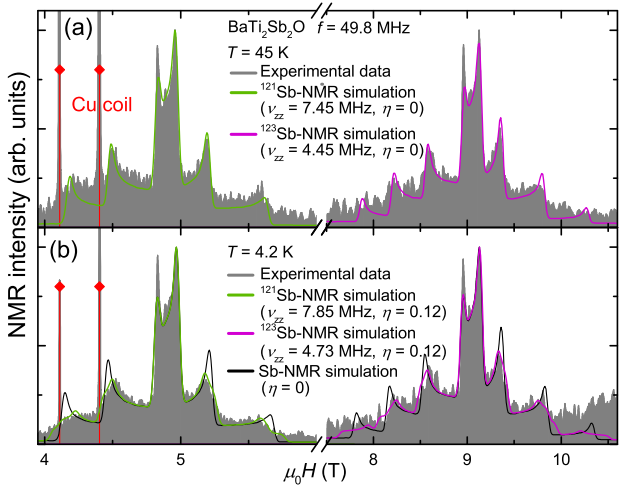}
\end{center}
\caption{(Color online) Field-swept $^{121/123}$Sb-NMR spectra measured at 45 K(a) and 4.2 K (b) measured at 49.8 MHz.
Simulations of NMR spectra calculated with the same NQR parameter as NQR spectra are also shown.
The measured NMR spectra can be consistently reproduced by the simulation, showing the validity of the NQR analysis.
For comparison, simulations of NMR spectra with $\eta = 0$ are shown at $T = 4.2$~K.}
\label{Fig.4}
\end{figure}

Reflecting the four-fold symmetry of the crystal structure, $\eta$ is zero at 45~K, while the NQR spectrum is gradually shifted  below 40 K.
$T$ dependence of the resonance frequencies arising from the $\pm$1/2 $\leftrightarrow$ $\pm$3/2 ($\nu_1$) and $\pm$3/2 $\leftrightarrow$ $\pm$5/2 ($\nu_2$) transition of $^{123}$Sb is shown in Fig.~\ref{Fig.3} (a). 
In the case of $I = 7/2$, NQR frequencies from each transition can be described as 
\begin{align*}
\nu_1 = \nu_{zz} \left(1 + \frac{109}{30}\eta^2\right), \text{and~} 
\nu_2 = 2\nu_{zz} \left(1 - \frac{17}{30}\eta^2\right)
\end{align*}
within the second-order perturbation of $\eta$ in $\mathcal{H}_Q$.
Above 40~K, the experimental result shows the relation $\nu_2/\nu_1 \simeq 2(1-\frac{21}{5}\eta^2) \simeq 2$ within the experimental error, but at 4.2~K, the result shows $\nu_2/\nu_1 \simeq 1.9$, which is evidence of finite $\eta$.
From the above resonance frequencies, we derived $T$ variation of $\eta$ and $\nu_{zz}$, which is shown in Fig.~\ref{Fig.3} (b) and the inset, respectively.
The $T$ dependence of $\nu_{zz}$ is consistent with that in the lattice parameters\cite{T.Yajima_JPSJ_2012}, indicative of the validity of the estimation.
The spectra below $T_{\rm A}$ can be interpreted by the change of $\nu_{zz}$ and the finite $\eta$ without an internal field appearing at the Sb site. 
This indicates the breaking of the in-plane four-fold symmetry at the Sb site at low temperatures. 
The $\eta$ changes continuously below 40~K and no clear hysteresis is observed, showing the transition at $T_{\rm A}$ being 2nd order.

\begin{figure}[!tb]
\vspace*{-10pt}
\begin{center}
\includegraphics[width=9cm,clip]{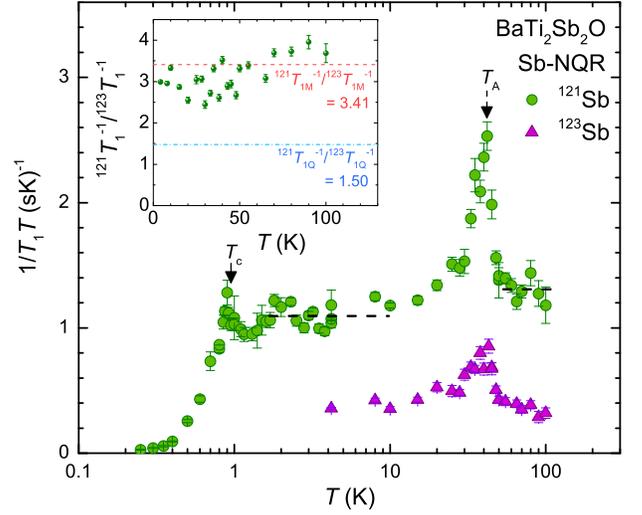}
\end{center}
\caption{(Color online) $T$ dependence of $1/T_1T$ measured with $^{121/123}$Sb-NQR.
1/$T_1T$ is constant above 60~K, suggestive of the conventional Fermi liquid (FL) state.
On cooling, 1/$T_1T$ shows a peak at $T_{\rm A} \sim$ 40~K.
On further cooling, 1/$T_1T$ recovers the FL state below 10~K with the slightly smaller values than those above 50~K. 
(Inset) $T$ dependence of the $^{121/123}$Sb isotopic ratio $^{121}T_1^{-1}$/$^{123}T_1^{-1}$. $^{121}T_{1}^{-1}/^{123}T_{1}^{-1} \sim 3$, indicating that the magnetic process is dominant in the whole temperature range.}
\label{Fig.5}
\end{figure}

Next, we focus on $T$ dependence of low-energy spin dynamics probed with $1/T_1$ measurements at the Sb site.
Figure~\ref{Fig.5} shows $T$ dependence of 1/$T_1T$ of $^{121/123}$Sb-NQR in BaTi$_2$Sb$_2$O.
Values of $T_1$ were derived by fitting the recovery data of $R(t) = 1- m(t)/m(\infty)$ to the theoretical NQR recovery curves of $I$ = 5/2 and 7/2.
Here $m(t)$ is the time dependence of spin-echo intensity at the peak of the NQR spectrum after saturation of nuclear magnetization, and $R(t)$ in whole measurement range could be fit with a single component of $T_1$.
1/$T_1T$ = const. (so-called ``Korringa'') behavior was observed above $\sim$ 50~K, suggestive of the Fermi liquid (FL) state.
On cooling, 1/$T_1T$ starts to increase at 50 K, where the resistivity shows a kink, and exhibits a peak at around 40~K, followed by the FL state again on further cooling.
Constant value of 1/$T_1T$ in a $T$ range between 1.5~K and 10~K is $\sim 1.1 \pm 0.03$, which is $\sim 83$\% of $1/T_1T$ value above 50~K, indicative of the decrease of  $N(E_F)$ by 9\% below $T_{\rm A}$, since 1/$T_1T$ in the FL state is proportional to square of density of states (DOS) around Fermi energy $N(E_F)$.

To investigate the origin of the phase transition, we measured 1/$T_1T$ at two Sb isotopes and estimate $1/T_1$ ratio between the $^{121/123}$Sb isotopes ($^{121}T_1^{-1}$/$^{123}T_1^{-1}$) as shown in the inset of Fig.~\ref{Fig.5}.
In general, an NMR/NQR spin lattice relaxation occurs through magnetic and/or electric-quadrupole channels. 
In the case of magnetic channel, fluctuations of local magnetic fields at a nuclear site cause magnetic relaxations between the nuclear spin levels of $\Delta m = \pm1$.
The magnetic relaxation rate is related to the gyromagnetic ratio $\gamma_n$ by $T_{1M}^{-1} \propto \gamma_n^2$.
Therefore, 
\begin{align*}
\frac{^{121}T_{1M}^{-1}}{^{123}T_{1M}^{-1}} = \left(\frac{10.189}{5.5175}\right)^2 = 3.41 
\end{align*}
is estimated from Tab. I.

On the other hand, fluctuations of EFG cause electric quadrupole relaxation between the nuclear spin levels of $\Delta m = \pm1$ and $\Delta m = \pm2$.
In this case, the electric-quadrupole relaxation rate is related to the quadrupole moment $Q$ by $T_{1Q}^{-1} \propto 3(2I+3)Q^2/[10(2I-1)I^2]$\cite{Y.Obata_JPSJ_1964}.
From obtained $\nu_{zz} \equiv 3e^2qQ/2I(2I-1)$, the ratio of $^{121}Q/^{123}Q$ can be estimated as,
\begin{align*}
\frac{^{121}Q}{^{123}Q} = \frac{\frac{2}{3}\cdot \frac{5}{2}(2\cdot \frac{5}{2}-1)\cdot ^{121}\nu_{zz}}{\frac{2}{3}\cdot \frac{7}{2}(2\cdot \frac{7}{2}-1)\cdot ^{123}\nu_{zz}} \simeq 0.80
\end{align*}
Then,
\begin{align*}
\frac{^{121}T_{1Q}^{-1}}{^{123}T_{1Q}^{-1}} = \frac{\frac{3(2\cdot \frac{5}{2}+3)}{10(2\cdot \frac{5}{2}-1)\left(\frac{5}{2}\right)^2}}{\frac{3(2\cdot \frac{7}{2}+3)}{10(2\cdot \frac{7}{2}-1)\left(\frac{7}{2}\right)^2}}\left(\frac{^{121}Q}{^{123}Q}\right)^2\simeq 2.34\times 0.80^2 = 1.50
\end{align*}
is calculated.

From the NQR measurements, the ratio of $^{121}T_{1}^{-1}/^{123}T_{1}^{-1}$ is $\sim 3$, indicating that the magnetic relaxation process is dominated in whole temperature range and magnetic fluctuations enhance toward $\simeq 40$~K.

Here we discuss the origin of the anomaly at $T_{\rm A}$ on the basis of experimental results reported so far.
As mentioned above, a CDW /SDW ordering accompanied with the anomaly in the lattice parameters was suggested from the resistivity and susceptibility measurements, although a tetragonality is maintained below $T_{\rm A}$\cite{T.Yajima_JPSJ_2012,T.C.Ozawa_JSSC_2000}. 
On the other hand, the $^{121/123}$Sb-NMR/NQR spectra at low temperatures revealed that the in-plane four-fold symmetry is broken at the Sb site below $T_{\rm A}\sim$ 40K without the internal field appearing at the Sb site. 
The absence of an internal field excludes the SDW ordering with an incommensurate correlation, since otherwise internal fields should appear at the Sb site, resulting in that resonance peaks, particularly those arising from the transition between $m = \pm 1/2$ and $\pm 3/2$, are split or broadened. 

Alternatively, the change of the NQR spectra below $T_{\rm A}$ can be understood by the occurrence of the commensurate CDW ordering.  
When a CDW transition occurs, there appear several Ti sites with different charge densities in most cases. 
However, the shift of the $^{121/123}$Sb NQR peak without splitting nor appreciable broadening below $T_{\rm A}$ gives a strong constraint and indicates that charge densities at the Ti sites should have a commensurate correlation, since there would be several Sb sites induced below $T_{\rm A}$ if a CDW ordering possesses an incommensurate correlation.    
The different charge densities at the Ti sites, e.g. the Ti configuration shown in Fig.~\ref{Fig.3}(b) breaks the in-plane four-fold symmetry at the Sb site although a tetragonality is maintained below $T_{\rm A}$.
The charge difference at the Ti site is considered to be small since the change of the DOS probed with $1/T_1T$ and the kink of the resistivity are very small below $T_{\rm A}$. 
However, the possibility of the magnetic ordering together with the CDW ordering at $T_{\rm A}$ would not be ruled out, since we cannot exclude commensurate magnetic ordering with a specific relationship between  magnetic correlations and ordered moment direction. (e.g. internal fields are canceled out at the Sb site when magnetic correlations are checkerboard $(\pi, \pi)$  and ordered moments direct to the $c$-axis, even if the off-diagonal hyperfine fields are taken into account.\cite{K.Kitagawa_JPSJ_2008, S.Kitagawa_PRB_2010})
To exclude the possibility of magnetic ordering, NMR/NQR measurements at the Ti site and/or neutron scattering measurements are crucial.
It should be noted that magnetic fluctuations are enhanced toward $T_{\rm A}$. 
In a well-known 2$H$-NbSe$_2$, where superconductivity ($T_{\rm c} \sim 7$ K) occurs below the CDW transition at $T_{\rm CDW} \sim$ 35 K, no anomaly was observed in $1/T_1$ at $T_{\rm CDW}$\cite{K.Ishida_JPSJ_1996}, however the similar anomaly of $1/T_1$ as in BaTi$_2$Sb$_2$O was observed in Lu$_5$Ir$_4$Si$_{10}$ at $T_{\rm CDW}$\cite{Y.Nakazawa_JPS_2003}. 
Since the relationship between charge and magnetic degree of freedom has not been well understood, BaTi$_2$Sb$_2$O might be one of the suitable system to investigate a correlation between charge and spin dynamics.

\begin{figure}[tb]
\vspace*{-10pt}
\begin{center}
\includegraphics[width=11cm,clip]{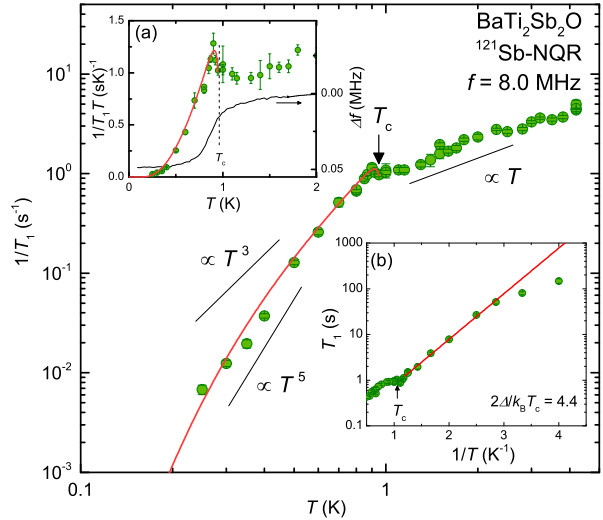}
\end{center}
\caption{(Color online) $T$ dependence of $1/T_1$ measured at 8.0~MHz. The inset (a) shows the $T$ dependence of $1/T_1T$ and a diamagnetic shielding signal measured by an NMR coil around $T_{\rm c}$, and the inset (b) shows the Arrhenius plot of $T_1$. As seen in the inset (a) and (b), $1/T_1$ shows a coherence peak just below $T_{\rm c} \simeq 0.95$~K and rapidly decreases at low temperatures.
The red curves are a calculation of $T$ dependence of $1/T_1$ based on an $s$-wave model with a finite gap. In the $s$-wave model, $2\Delta/k_{\rm B}T_{\rm c} = 4.4$ and $\delta/\Delta = 0.5$ are adopted, where $\delta / \Delta$ are the broadening parameter of the SC gap.
$1/T_1$ results in the SC state is consistently reproduced with the $s$-wave model. }
\label{Fig.6}
\end{figure}

Next, we discuss the $1/T_1$ in the SC state.
$1/T_1$ of $^{121}$Sb-NQR slightly decreases below 1.5 K, where small Meissner signal appears, but $1/T_1$ shows a tiny coherence peak just below $T_{\rm c} \simeq 0.95$~K, where the sharp Meissner signal is observed, and then rapidly decreases at low temperatures as shown in Fig.~\ref{Fig.6}.
$T$ dependence of $1/T_1$ far below $T_{\rm c}$ is much steeper than $T^3$ dependence, but $T_1$ follows an exponential $T$ dependence down to 0.3~K, as shown by an Arrhenius plot in the inset (b). 
These are in sharp contrast with $T$ dependence in cuprate\cite{K.Ishida_PhysicaC_1991} and iron-based superconductors\cite{K.Ishida_JPSJ_2009}.
From the slope of the plot, the magnitude of the SC gap is estimated to be $2\Delta/k_{\rm B}T_{\rm c} = 4.4$, and actually the observed $1/T_1$ in the SC state can be fit consistently by an $s$-wave full gap model with $2\Delta/k_{\rm B}T_{\rm c} = 4.4$ and $\delta/\Delta = 0.5$, where $\delta$ is the broadening parameter of the singularity in the SC DOS.
Absence of the residual DOS in the SC state is also consistent with an $s$-wave model, since residual DOS suggested by the Korringa behavior far below $T_{\rm c}$ is easily introduced by disorder and/or a tiny amount of impurities in unconventional superconductors\cite{Y.Nakai_PRB_2010,K.Ishida_PRB_1997}.
Since the present NQR measurements were done in an ``early-stage'' polycrystalline sample, full-gap $s$-wave state would be the most possible SC gap state in BaTi$_2$Sb$_2$O.

In summary, the NQR asymmetric parameter $\eta$ becomes finite below $T_{\rm A} \simeq$ 40~K, indicative of the breaking of the in-plane four-fold symmetry at the Sb site without internal field appearing at the Sb site. The variation of the NQR spectra below $T_{\rm A}$ can be understood by the occurrence of the commensurate CDW transition.
$1/T_1$ below $T_{\rm A}$ shows a further anomaly below $T_{\rm c}$ due to the opening of the SC gap, indicative of the coexistence of superconductivity and the anomaly occurring at $T_A$.
In the SC state, $1/T_1$ shows a coherence peak just below $T_{\rm c}$ and exponentially decreases at low temperatures, which strongly suggests that the SC symmetry of BaTi$_2$Sb$_2$O is an $s$-wave with finite SC gap.

The authors thank to Y. Nakai, K. Kitagawa, S. Yonezawa, and Y. Maeno for experimental support and valuable discussions. 
The authors also grateful to T. Tohyama for fruitful discussion. 
This work was partially supported by Kyoto Univ. LTM center, the ``Heavy Electrons'' Grant-in-Aid for Scientific Research on Innovative Areas  (No. 20102006) from The Ministry of Education, Culture, Sports, Science, and Technology (MEXT) of Japan, a Grant-in-Aid for the Global COE Program ``The Next Generation of Physics, Spun from Universality and Emergence'' from MEXT of Japan, a grant-in-aid for Scientific Research from Japan Society for Promotion of Science (JSPS), KAKENHI (S and A) (No. 20224008 and No. 23244075) and FIRST program from JSPS. One of the authors (SK) is financially supported by a JSPS Research Fellowship.


%

\end{document}